\documentclass[11pt]{article}

\usepackage[margin=1in]{geometry}
\usepackage{graphicx}
\usepackage{booktabs}
\usepackage{tabularx}
\usepackage{array}
\usepackage{amsmath}
\usepackage{url}

\graphicspath{{figures/}}
\title{CFDTwin: An open-source GUI and Python toolkit for POD-NN surrogate modeling of ANSYS Fluent simulations}

\author{Daniel Curl and Han Hu\thanks{Corresponding author: \texttt{hanhu@uark.edu}}\\
Department of Mechanical Engineering, University of Arkansas, Fayetteville, AR 72701, USA}

\date{}

\begin{document}

\maketitle

\begin{abstract}
High-fidelity computational fluid dynamics (CFD) is widely used for thermal-fluid design, but repeated CFD solves remain expensive for design optimization, uncertainty analysis, and digital-twin workflows. Recently, our team has demonstrated that a proper orthogonal decomposition and neural-network (POD-NN) surrogate can predict two-dimensional thermal fields in an electronics-cooling cold plate with large inference speedups while preserving physically interpretable modal structure. Reproducing and extending such workflows, however, typically requires custom scripts for parameter sampling, Fluent automation, data extraction, reduced-order model construction, neural-network training, validation, and prediction. This paper introduces CFDTwin, an open-source Python package and optional desktop graphical user interface (GUI) that packages these steps into a reusable workflow for ANSYS Fluent simulations. CFDTwin allows users to define simulation inputs and output quantities, generate design-of-experiments samples, run and resume Fluent batch simulations, train POD-NN surrogate models for scalar, surface-field, and cell-zone outputs, inspect validation metrics, and evaluate trained models at new design points without re-running Fluent. The same workflow is exposed through a scriptable Python API and a GUI, supporting reproducible studies, user-facing model validation, and automated design exploration. CFDTwin extends the prior POD-NN modeling study from a case-specific research implementation to a reusable research-software platform for CFD surrogate modeling and digital-twin development.
\end{abstract}


\section{Introduction}

Thermal-fluid design increasingly depends on high-fidelity CFD because modern systems often involve coupled geometry, material, flow, and heat-transfer interactions that are difficult to capture with closed-form models alone. In electronics cooling, for example, rising power densities motivate rapid evaluation of cold-plate geometries, coolant flow rates, heat loads, and boundary conditions. CFD can resolve the relevant flow and thermal fields, but repeated CFD solves remain too expensive for workflows that require hundreds or thousands of candidate evaluations, such as design optimization, sensitivity analysis, uncertainty quantification, and real-time digital twins.

Digital-twin concepts are increasingly relevant to electronics cooling because thermal-management systems must be evaluated during both design and operation. Industrial digital twins integrate models, data, and decision workflows to represent and predict the behavior of physical systems \cite{tao2019digitaltwin}. In computational mechanics, reduced-order models, surrogate models, optimization, and machine learning are commonly identified as enabling technologies for making digital twins fast enough for repeated evaluation \cite{stavroulakis2022review}. Electronics-cooling applications have begun to adopt this view; for example, recent work has demonstrated a digital-twin framework for real-time optimization of a microchannel heat sink using model feedback and learning-based decision making \cite{xue2026microchannel}. These studies motivate software that can convert high-fidelity thermal-fluid simulations into fast, reusable surrogate models.

Reduced-order and data-driven surrogate models address this bottleneck by learning an approximate mapping from design or operating parameters to simulation outputs, and recent reviews have identified machine-learning-assisted CFD as an active path toward faster simulation workflows \cite{panchigar2022mlcfd}. Proper orthogonal decomposition (POD) is especially attractive for field prediction because it represents high-dimensional CFD outputs using a small set of dominant spatial modes and modal coefficients \cite{holmes1997pod}. Neural networks can then be trained to map input parameters to those coefficients, allowing rapid reconstruction of scalar, surface-field, or volumetric outputs; this non-intrusive POD-NN strategy has been demonstrated for nonlinear reduced-order modeling and several CFD field-prediction settings \cite{hesthaven2018nonintrusive,macraild2024aneurysm,ganti2020multiphase}. In our past work, we have demonstrated a POD-NN surrogate for two-dimensional thermal-field prediction in a liquid-cooled dual-chip cold plate and showed that the learned modal structure could be connected to physically meaningful thermal features \cite{curl2026aitf}. Despite these proven successes, there exists a practical barrier for AI-based surrogate modeling, i.e., the workflow accessibility of the model. A typical CFD surrogate study requires users to connect several tasks that are often implemented separately, including selecting input parameters from a CFD case, sampling a bounded design space, running a batch of simulations, extracting consistent outputs, training reduced-order and machine-learning models, storing model artifacts, validating predictions, and applying the trained surrogate to new points. When these steps are handled by custom scripts, the resulting workflow can be difficult for new users to reproduce, difficult for collaborators to audit, and cumbersome to adapt to new cases.

CFDTwin was developed to address this usability and reproducibility gap. The software packages the POD-NN surrogate workflow into an open-source Python package and a wizard-based desktop GUI for ANSYS Fluent simulations. Users can either step through the workflow interactively or call the same pipeline through a Python API. This paper describes the motivation, design, implementation, and use of CFDTwin, with emphasis on how the software extends the prior POD-NN research from a demonstrated modeling framework to a reusable tool for CFD surrogate modeling, design exploration, and digital-twin workflows.

The main contributions of this paper are:
\begin{enumerate}
    \item An open-source software package that automates the end-to-end workflow for building neural-network surrogate models from ANSYS Fluent simulations.
    \item A GUI that lowers the practical barrier to defining inputs, generating DOE samples, running simulations, training models, validating predictions, and analyzing surrogate results.
    \item A scriptable Python API that exposes the same workflow for reproducible studies, parameter sweeps, and integration with optimization routines.
    \item A project-based storage structure that records case references, input/output specifications, DOE samples, simulation outputs, model artifacts, and validation metadata.
    \item Documentation and a video tutorial that support adoption by users who may not be familiar with custom CFD automation or machine-learning workflows.
\end{enumerate}

Different from our past study focused on demonstrating a POD-NN surrogate for thermal fields prediction with high speedup, controlled error sources, and physically interpretable modal behavior \cite{curl2026aitf}, the present work addresses the next practical question, i.e., how can the surrogate-modeling workflow be made usable and reproducible for broader CFD studies? Rather than introducing a new reduced-order modeling theory, CFDTwin packages the workflow into software that can be applied to user-provided Fluent cases. The software handles project setup, DOE generation, Fluent batch execution, output extraction, model training, validation, and prediction. The contribution is therefore a reusable research-software implementation of a POD-NN surrogate-modeling workflow, with the earlier article serving as the primary methodological and validation reference.

\section{Software Description}

\subsection{Software Overview}

CFDTwin is an open-source Python package for constructing neural-network surrogate models from ANSYS Fluent simulations. The current implementation supports Python 3.10 or newer and requires a working ANSYS Fluent installation. The software is distributed as the \texttt{cfdtwin} Python package and is released under the MIT license. The source code is hosted in the UARK-NED3/CFDTwin GitHub repository, the version described in this manuscript is archived through Zenodo with DOI 10.5281/zenodo.20249626, and user documentation is provided through the project website \cite{cfdtwin2026github,cfdtwin2026zenodo,cfdtwin2026docs}.

The software provides two interfaces to the same workflow. The first is a lightweight Python API installed with \texttt{pip install cfdtwin} and centered on a \texttt{Project} object. This API enables reproducible scripts, automated parameter sweeps, and later coupling with optimization or control algorithms. The second is an optional wizard-based desktop GUI installed with \texttt{pip install "cfdtwin[gui]"} and launched with the \texttt{cfdtwin-gui} command. The GUI organizes the workflow into five stages: Setup, DOE, Simulate, Train, and Analyze. This structure is designed to mirror the sequence of decisions required in a surrogate-model study while preventing users from advancing to later stages before prerequisites are complete.

The core workflow begins when a user creates or opens a CFDTwin project and points it to a Fluent case file. The user then defines input parameters, such as boundary-condition values or Fluent input parameters, and specifies outputs to extract from each simulation. Outputs may include scalar report definitions, surface fields, or cell-zone fields. CFDTwin then generates a design of experiments using Latin hypercube sampling or a factorial design, executes the required Fluent simulations, stores each simulation result, trains one surrogate model per configured output, and evaluates trained models at new input points.

This structure is intended to preserve the scientific logic of a CFD surrogate workflow. The user must explicitly state the input design space, the output quantities of interest, and the sampling plan before model training begins. This design helps prevent the surrogate from being treated as a black-box predictor detached from the underlying CFD setup. It also makes the project folder a record of the assumptions and artifacts used to construct the surrogate.

\subsection{Software Architecture and Workflow}

CFDTwin is organized around a project directory that acts as the persistent record of a surrogate-modeling study. A project stores the case-file reference, input definitions, output definitions, DOE samples, extracted simulation data, trained model files, loss curves, metrics, and metadata. This design allows interrupted simulation batches to be resumed, trained models to be inspected after the fact, and predictions to be reproduced without reconstructing the original session state.

Table~\ref{tab:project-layout} summarizes the main files and directories created within a CFDTwin project. The configuration files define the study, the DOE file records the sampled design space, the \texttt{dataset/} directory stores the CFD-derived training data, and the \texttt{models/} and \texttt{results/} directories store the trained surrogate and validation outputs. Because the GUI and Python API read and write the same project structure, a project created interactively can be inspected, reproduced, or extended through scripts.

\begin{table}[htbp]
\centering
\caption{Main files and directories in a CFDTwin project.}
\label{tab:project-layout}
\small
\begin{tabularx}{\textwidth}{>{\ttfamily}p{0.25\textwidth}p{0.25\textwidth}X}
\toprule
\normalfont File or directory & Role in the workflow & Typical contents \\
\midrule
project\_info.json & Project-level record & Project name, case-file path, creation metadata, and Fluent/session settings \\
model\_setup.json & Surrogate-model specification & Input definitions, output definitions, DOE settings, and training configuration \\
doe\_samples.csv & Design-of-experiments table & Sampled input combinations used for Fluent simulations \\
dataset/sim\_*.npz & Per-simulation output data & Extracted scalar, surface-field, or cell-zone outputs for completed Fluent runs \\
dataset/metadata.json & Dataset summary & Simulation status, output dimensions, extraction settings, and dataset-level metadata \\
models/ & Trained surrogate artifacts & Neural-network weights, POD modes, scalers, model metadata, and loss histories \\
results/ & Analysis outputs & Validation metrics, prediction summaries, comparison data, and generated plots \\
logs/ & Execution trace & Simulation, training, warning, and error logs \\
\bottomrule
\end{tabularx}
\end{table}

The public API is centered on the \texttt{cfdtwin.Project} class. A typical script creates a project, sets the case file, declares inputs and outputs, generates a DOE, connects to Fluent, runs simulations, trains a model, and predicts at new design points. For example, the project documentation demonstrates a mixing-elbow case in which inlet velocity magnitudes are varied, outlet temperature is extracted, and a POD-NN surrogate is trained from the resulting simulation dataset \cite{cfdtwin2026docs}.

The software separates the workflow into modular components. The DOE module generates Latin hypercube or factorial samples within user-defined parameter ranges, following the broad use of space-filling input selection methods for computer experiments \cite{mckay1979lhs}. The Fluent interface launches a Fluent session, loads the configured case, applies each sample's input values, runs the solver, and extracts the requested outputs. The simulation module stores per-sample results in compressed NumPy files. The training module detects output type, loads the dataset, partitions training and testing samples, applies POD reduction when appropriate, trains neural-network regressors, and writes model artifacts and metrics. The results module provides structured return objects for simulation, training, and prediction summaries.

\begin{figure}[htbp]
    \centering
    \includegraphics[width=\textwidth]{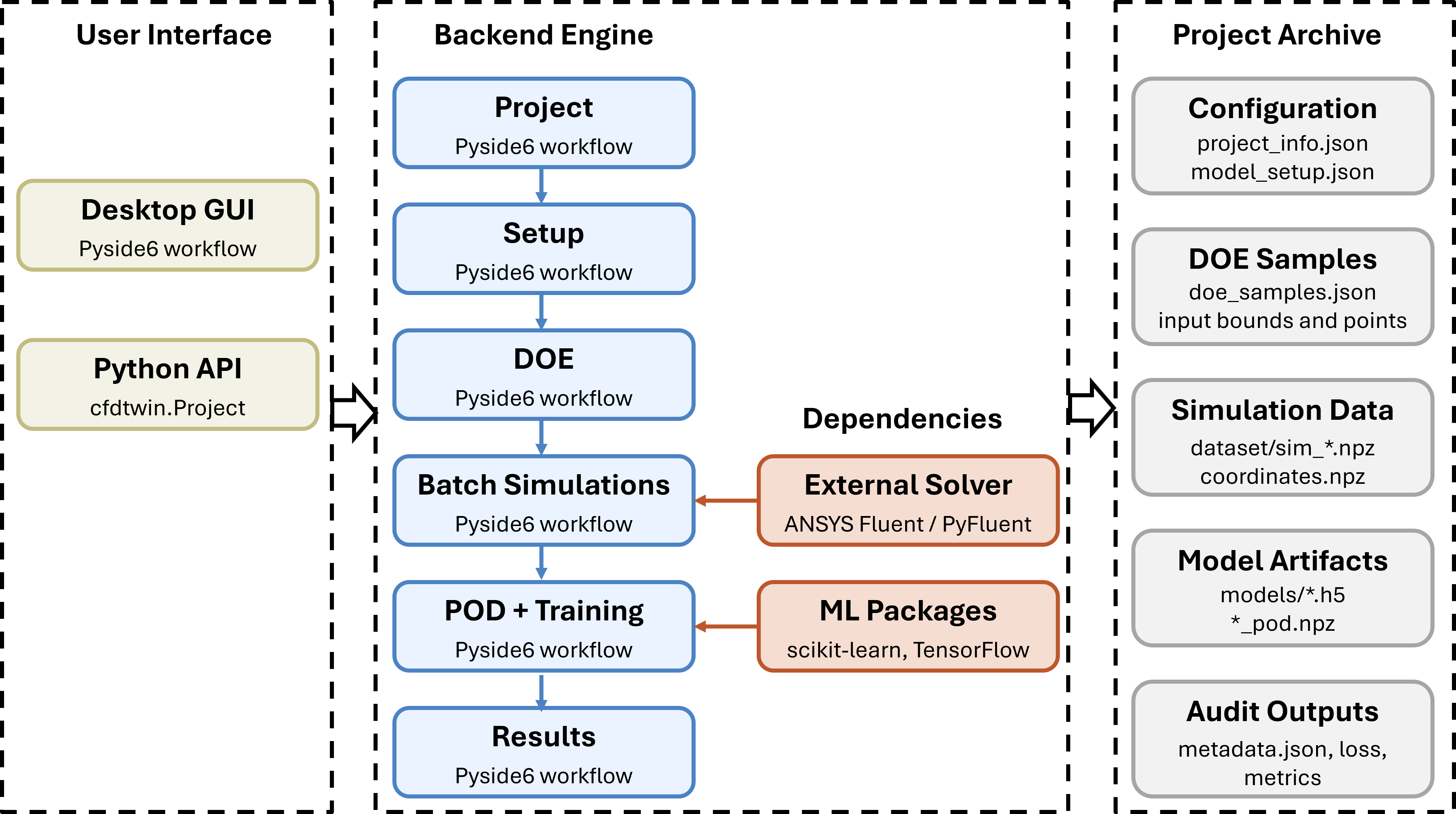}
    \caption{Project architecture and workflow for surrogate modeling using CFDTwin.}
    \label{fig:workflow}
\end{figure}

Figure~\ref{fig:workflow} summarizes the resulting architecture and workflow. The left side shows that users can enter the workflow either through the desktop GUI or through the \texttt{cfdtwin.Project} API. The central backend engine represents the same ordered workflow used by both interfaces: project creation, setup, DOE generation, batch simulation, POD and neural-network training, and results analysis. The right side shows the project archive produced by the workflow, including configuration files, DOE samples, simulation data, trained model artifacts, and audit outputs. This organization is central to the software contribution because the GUI, API, and archived project state remain consistent with one another.

The GUI uses the same underlying project structure and computational modules. This is important because the GUI is not a separate demonstration path; it is a user-facing layer over the same reproducible workflow available through scripts. A user can therefore begin with the GUI to learn the workflow and later move to the API for automation, or use scripts to generate a project that can be inspected through the GUI.

\subsection{Model Construction}

CFDTwin constructs surrogate models from CFD output data generated over a sampled input design space. Let \(x\) denote the vector of user-defined simulation inputs and \(y(x)\) denote a scalar, surface-field, or cell-zone output extracted from Fluent. For scalar report definitions, CFDTwin trains a neural-network regression model directly from input parameters to scalar outputs. For field outputs, CFDTwin uses a reduced-order representation so that the neural network predicts a low-dimensional coefficient vector rather than every spatial point directly.

For field data, the extracted simulation outputs are assembled into a snapshot matrix after all successful simulations are complete. POD is then used to obtain dominant modes that represent the principal spatial variations in the training data. The retained POD coefficients become the regression targets for a neural network. During prediction, the trained neural network maps a new input vector to POD coefficients, and the predicted field is reconstructed from the retained modes and mean field. This approach follows the POD-NN framework established in the prior cold-plate study \cite{curl2026aitf} and is consistent with earlier non-intrusive POD-NN and data-driven CFD surrogate modeling work \cite{hesthaven2018nonintrusive,macraild2024aneurysm,ganti2020multiphase}, while CFDTwin generalizes the workflow into a reusable software pipeline.

The software stores information needed to audit each trained model, including the output location, field variable, output type, number of spatial points, model architecture, training and testing metrics, train-test split, training date, number of POD modes, and variance explained when POD is used. This metadata is intended to make trained models more transparent than standalone neural-network files and to support comparison among training runs.

\subsection{GUI Workflow}

The CFDTwin GUI was developed to make the surrogate-modeling workflow accessible to users who may not want to write custom automation scripts. The GUI is organized as a sequence of workflow pages. In the Setup page, the user selects a Fluent case file, configures Fluent options, and defines inputs and outputs. In the DOE page, the user generates samples using Latin hypercube or factorial sampling. In the Simulate page, the user launches batch Fluent simulations and monitors progress. In the Train page, the user trains surrogate models and inspects loss curves. In the Analyze page, the user views metrics, generates predictions, and compares surrogate predictions with Fluent results when available.

This design has two benefits for research use. First, it reduces onboarding time for new users because the interface exposes the required modeling decisions in the order they must be made. Second, it improves reproducibility because the GUI writes the same project files used by the API. The workflow is therefore not hidden inside a manual sequence of clicks; it is reflected in project artifacts that can be inspected, reused, or transferred to scripted workflows.

\begin{figure}[htbp]
    \centering
    \includegraphics[width=\textwidth]{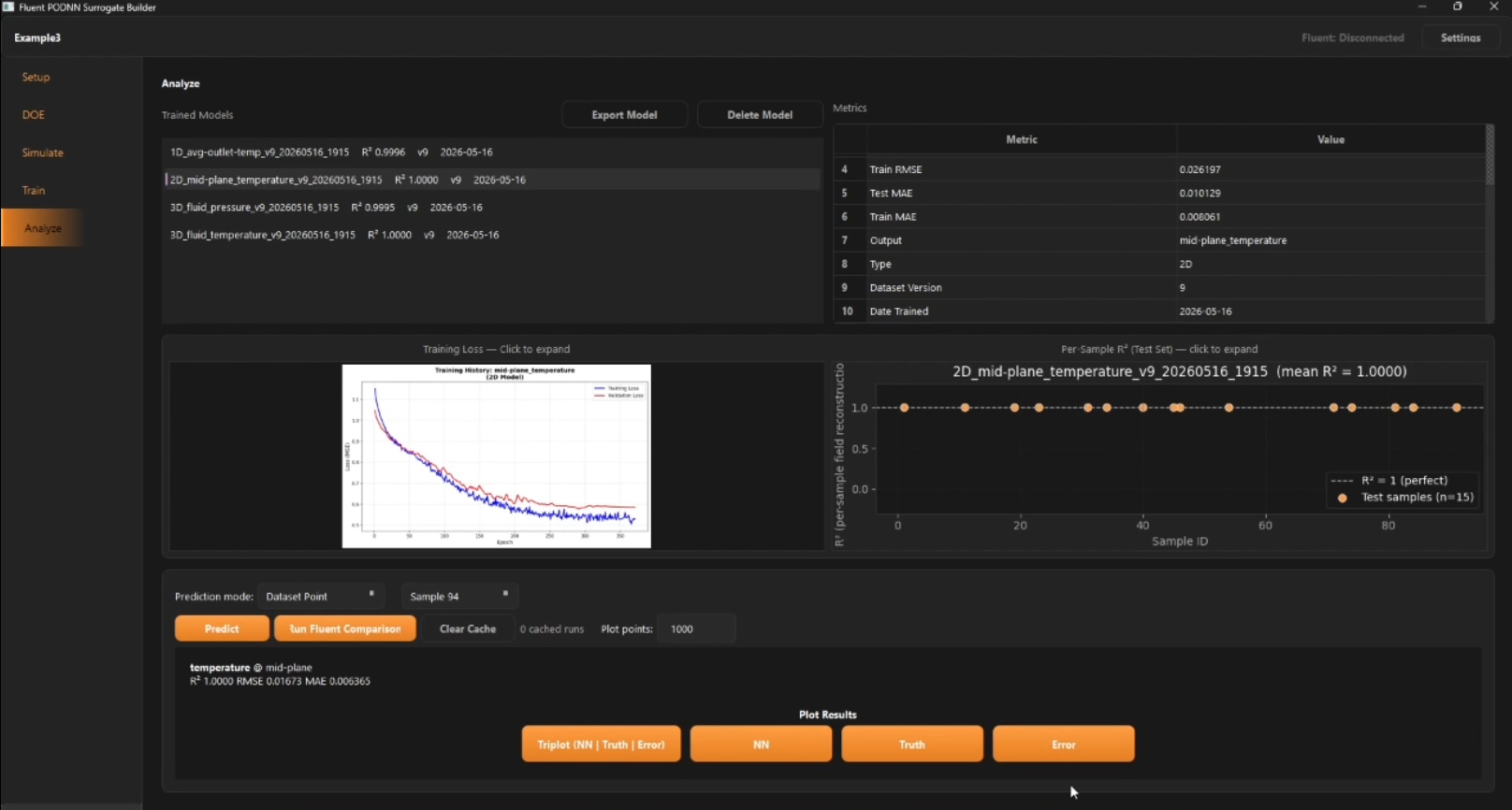}
    \caption{Snapshot of the graphical user interface of CFDTwin showing the Analyze page.}
    \label{fig:gui}
\end{figure}

Figure~\ref{fig:gui} shows the Analyze page of the GUI after surrogate models have been trained. The model list provides a compact record of available trained outputs, including scalar, surface-field, and cell-zone models. The metrics table reports training and testing errors for the selected model, while the loss-curve and per-sample test-set plots provide quick checks of model convergence and prediction consistency. The lower panel supports prediction and Fluent comparison for selected dataset points, including controls for plotting neural-network predictions, Fluent truth fields, and error fields. This page illustrates the practical role of the GUI: it turns model validation and prediction into a visible workflow rather than leaving those checks buried in scripts or output files.

\subsection{Python API Workflow}

The Python API supports the same workflow in a compact script. A minimal workflow has the following structure:

\begin{verbatim}
import cfdtwin

project = cfdtwin.Project.create("./my_study", name="study1")
project.set_case_file("mixing_elbow.cas.h5")

project.set_inputs({
    "cold-inlet|momentum > velocity_magnitude": (0.2, 0.6),
    "hot-inlet|momentum > velocity_magnitude":  (0.4, 1.2),
})

project.set_outputs([
    {"name": "outlet", "category": "Surface",
     "field_variables": ["temperature"]},
])

project.generate_doe(n=20, method="lhs")
project.connect_fluent(precision="single")
project.run_simulations(verbose=True)
result = project.train(model_name="run1")

pred = project.predict("run1", {
    "cold-inlet|momentum > velocity_magnitude": 0.4,
    "hot-inlet|momentum > velocity_magnitude":  0.8,
})
\end{verbatim}

This API is useful for users who want to automate studies after the workflow has been established. It also provides a practical path to design optimization. Once a model has been trained and validated, external optimization routines can call \texttt{project.predict(...)} repeatedly to evaluate candidate designs without launching Fluent. In this way, CFDTwin separates the expensive offline CFD sampling stage from the rapid online surrogate-evaluation stage.

\section{Design Optimization and Digital-Twin Use}

CFDTwin is designed to support downstream design optimization and digital-twin studies by separating expensive CFD data generation from rapid surrogate evaluation. In a typical offline-online workflow, users first define a design space, sample that space with CFD, train and validate a surrogate, and then use the trained model for repeated prediction. Once the surrogate is validated within the sampled design space, it can be coupled to optimizers, sensitivity-analysis routines, control algorithms, or graphical exploration tools.

\begin{figure}[htbp]
    \centering
    \includegraphics[width=\textwidth]{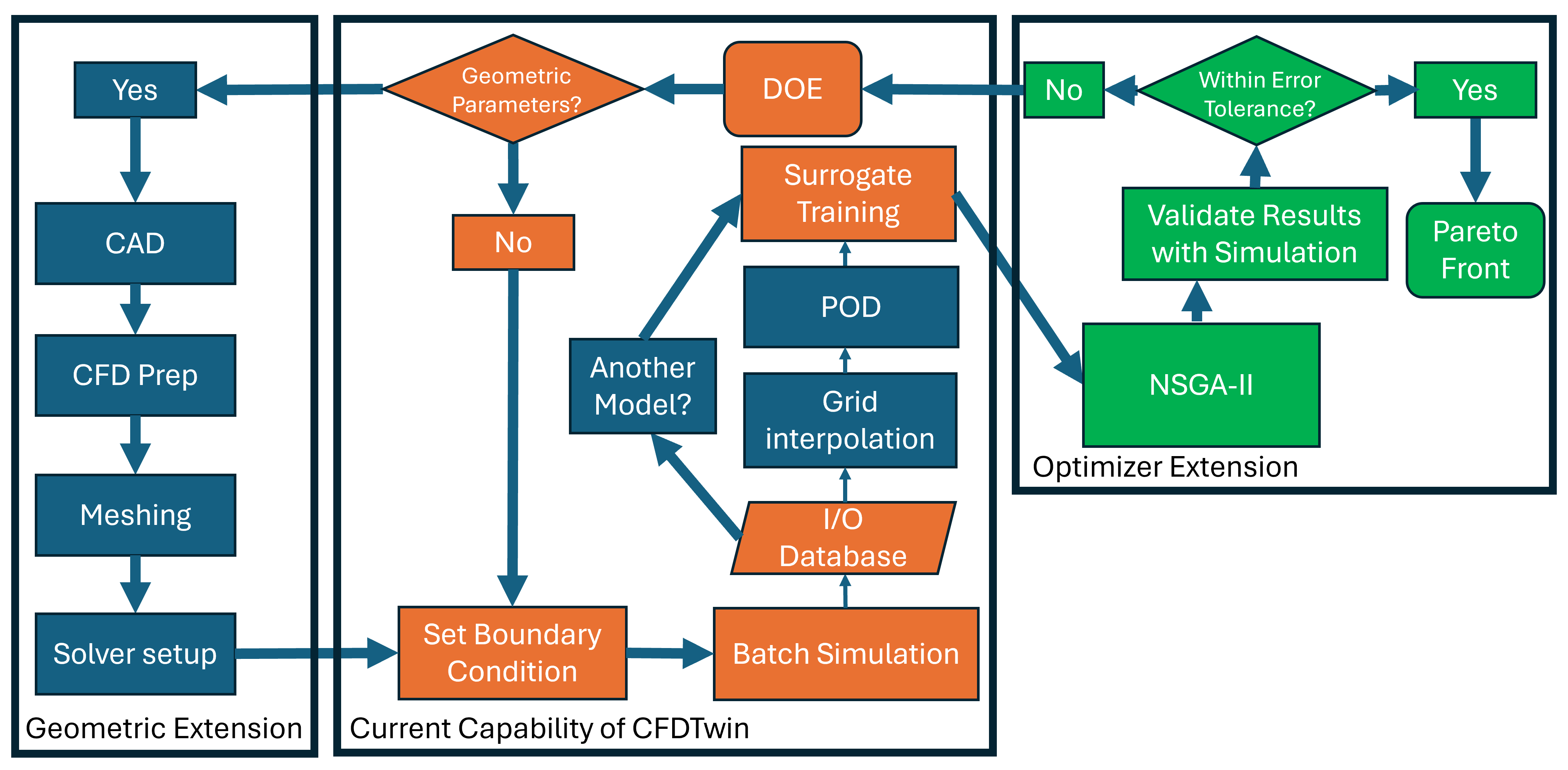}
    \caption{Design exploration using CFDTwin with geometric generation and optimization extensions.}
    \label{fig:design}
\end{figure}

Figure~\ref{fig:design} illustrates how the present CFDTwin workflow supports design exploration when combined with geometric generation and optimization extensions. The central portion of the figure represents the current CFDTwin capability: users define boundary conditions, run a DOE, execute batch simulations, train a POD-NN surrogate, and validate the surrogate against simulation results. The left branch shows a geometric extension in which CAD generation, CFD preprocessing, meshing, and solver setup automate the creation of new simulation-ready designs. The right branch shows an optimization extension in which surrogate predictions are used to evaluate candidate designs, check whether the error is within tolerance, and continue toward Pareto-front or NSGA-II-style optimization when the surrogate is sufficiently accurate. This figure clarifies the intended role of CFDTwin in a larger design loop: CFDTwin supplies the reusable surrogate-modeling core, while external geometry and optimization tools expand the workflow toward automated design exploration.

The software contribution is the construction and evaluation of the surrogate model. CFDTwin does not remove the need for CFD-quality checks, mesh verification, physically meaningful input bounds, or validation against held-out simulations. In particular, predictions outside the sampled input range are extrapolations and are not reliable unless separately validated. These limitations are central to data-driven CFD surrogate modeling.

In this workflow, a surrogate-assisted sweep or optimizer searches over inlet velocities or thermal boundary conditions to minimize maximum temperature, thermal resistance, pressure drop, or a weighted multi-objective score. CFDTwin does not guarantee an optimal design by itself; its role is to provide rapid evaluation of candidate designs after the surrogate has been trained and validated.

\section{Limitations and Future Development}

CFDTwin inherits the limitations of both CFD and data-driven surrogate modeling. The accuracy of any trained model depends on the fidelity of the original Fluent simulations, the quality of the mesh, the appropriateness of boundary conditions and solver settings, the coverage of the training design space, and the representativeness of the extracted outputs. CFDTwin automates the workflow but does not validate the underlying CFD model by itself.

The software currently targets ANSYS Fluent workflows and therefore requires a working Fluent installation. This dependency is appropriate for users whose high-fidelity simulations are already built in Fluent, but it limits immediate portability to open-source solvers or other commercial CFD packages. Future versions could generalize the simulation backend while preserving the project, DOE, training, and prediction architecture.

The current package is best suited to interpolation within a bounded design space. Predictions outside the training domain require caution. Future development could include uncertainty quantification, active learning, adaptive sampling, optimization modules, and automatic warnings for extrapolative prediction requests.

\section{Conclusions}

This paper introduced CFDTwin, an open-source GUI and Python toolkit for building POD-NN surrogate models from ANSYS Fluent simulations. CFDTwin packages the surrogate-modeling workflow into a project-based software system that supports input definition, DOE generation, batch simulation, model training, validation, prediction, and analysis. The software extends the prior POD-NN cold-plate study by turning a research modeling workflow into a reusable tool that can support CFD surrogate modeling, design exploration, and digital-twin development.

The main value of CFDTwin is practical reproducibility. It reduces the need for case-specific scripting, exposes the workflow through both GUI and API interfaces, stores model artifacts and metadata, and provides documentation and a video tutorial for new users. With the \texttt{v0.2.0} release archived on Zenodo, the software is citable and reproducible as a fixed research artifact.

\section*{Acknowledgments}

This study was supported by the National Science Foundation, United States, under Grant No. OIA- 2429580 and Grant No. TI-2431969.

\section*{Data and Software Availability}

The CFDTwin source code is available at \url{https://github.com/UARK-NED3/CFDTwin} under the MIT license. Documentation is available at \url{https://uark-ned3.github.io/CFDTwin/}. The version described in this manuscript is CFDTwin \texttt{v0.2.0}, archived on Zenodo at \url{https://zenodo.org/records/20249626} with DOI 10.5281/zenodo.20249626. The corresponding release tag resolves to commit \texttt{1f215326db5fc66d6dfbe594\allowbreak2633c13f5e199278}.

\bibliographystyle{unsrt}
\bibliography{references}

\end{document}